# Improving social media for democratic discourse

Authors in alphabetical order:
Fan Cheng[1], Amirhossein Farzmahdi[1], Pinyuan Feng[1], Kedar Garzón Gupta[1], Trenton Jerde[2*], Nikolaus Kriegeskorte[1*], Zi Qi Liow[1*], Akihito Maruya[1], Savannah Smith[1], Patrick Stinson[1], JohnMark Taylor[1]

[1]Columbia University, New York, USA
[2]Independent Researcher, New York, USA
*corresponding authors

## Executive summary

Designing social media platforms for democratic discourse is among the most important challenges of our time. Meeting this challenge will require ideas from people with a broad range of perspectives, including many who are not experts on social media. This white paper makes a start by assembling a wide range of ideas for improving social media and connecting them to relevant literature across multiple disciplines.

Social media are widely seen as causing or enabling political polarization, misinformation, disinformation, social isolation, and loss of civility. However, these harms are not unavoidable consequences of human nature or of being connected through technology. Rather, they reflect the design choices and incentives of the few platforms that dominate today. Today's platforms are optimized for engagement because it drives advertising revenue, and this objective is inherently in tension with the goals of democratic discourse. Beyond small experiments, designing social media for democratic discourse has not been seriously attempted. Designed to support democratic discourse, social media might yet deliver on their original promise of connecting people, broadening participation in political discourse, creating truly accessible national and global public spheres, and enhancing our collective intelligence.

To expand our practical imagination beyond the engagement paradigm, this white paper proposes a modular set of "mechanisms" that could be implemented on social media platforms. Each mechanism is presented as a one-page table, with a short description, bullet lists for its promise and challenges, references to related mechanisms in this white paper, and references to the literature. The collection is intended both as an entry point for discussion, with pointers to relevant literature, and as a menu of implementable design options. Our overarching goal is to catalyze broad discussion among the general public, journalists, and experts in multiple fields. The ideas may also inspire more technical, cross-disciplinary work in technology, engineering, the sciences, and the humanities, toward a digital public sphere that is built for democratic discourse, collective deliberation, and collective intelligence. Below is a brief overview of the proposed mechanisms (bold) for improving social media for democratic discourse.

### (1) Shift feedback and incentives from emotional engagement to epistemic value.

**Cognicons** are one-click cognitive reactions (analogous to emoticons but reflective of rational judgment) meant to amplify thoughtful and informative contributions rather than emotionally provocative ones. A related proposal is **probability ratings**, which let users annotate factual claims on a probability scale from 0 to 1 to enable decentralized fact-checking. The **number-free feed** would hide public like counts, share counts, and follower totals, encouraging users to assess content on its own merits rather than following perceived majority opinion. More broadly, the paper explores **game-aware design for democratic discourse**, where platforms can determine what content should receive amplification. Mechanisms such as cognicons, probability ratings, a **discourse score**, and **milestone badges for high-quality participation** could reward credible sourcing, constructive argument, cross-ideological engagement, and public belief revision.

### (2) Increase individual and community control, transparency, and participation.

**Natural language feed algorithm tuning** would let users give direct instructions to their recommender systems so that feeds reflect people's conscious goals, not just their habitual engagement. Similarly, **open-source forkable recommenders** would allow users to select among audited ranking algorithms, modify their settings, program their own algorithms, and compare feeds from alternative algorithms side by side. At the individual level, the **content-engagement summary and feed recalibration** mechanism would visualize what content users have engaged with, and which topics or perspectives may be missing. At the collective level, **communal feed curation** would let communities design and publish feeds based on shared values, while the **daily algorithm ballot** would allow users to vote on temporary feed priorities – for example, boosting local news. **Stream networks** are decentralized networks of collective intelligence, where streams of cultural items are created and curated by people, with each stream drawing selectively from multiple other streams using algorithms that are entirely controlled by the participants.

### (3) Expose the plurality of perspectives and engage across differences.

**Stance-based search** would make it easier to find arguments for, against, or neutral toward a particular position. A **devil's advocate button** would provide high-quality alternative viewpoints through retrieved posts, AI-generated counterarguments, or opt-in human advocates. **Virtual travel through geographic location filtering** would allow users to explore local conversations in other regions, while the **cross-lingual bridge layer** would make posts in other languages routinely accessible through automatic translation, confidence indicators, and community corrections. For developing events and complex discussions, **live opinion mapping** would organize posts into an interactive multi-perspective visualization. **Interactive bubble-chart opinion clustering** would use axes to visualize individual and collective stances from social media posts, fostering discussions across ideological clusters. **Timeline generation for controversial societal issues** would present the historical development of events and claims, while the **conversation graph for meaningful discourse** would visually label replies as supporting points, counterarguments, or clarifying questions. **Population-representative feeds** would match the visibility of opinions to their prevalence among the user base, amplifying widely held but under-expressed opinions and exposing in-groups to a larger range of opinions. To facilitate constructive interaction across ideological divides, **MatchAcross** would connect users across ideological divides, using social proximity (e.g., friends of friends) as a bridge.

### (4) Strengthen accuracy, civility, fairness, and evidentiary infrastructure.

**Collaborative fact-checking** would allow users to work together to investigate claims and provide context to content. **AI fact-checking** could work alongside it to flag claims and attach expert-reviewed sources directly to posts, with transparency about source and model selection. An **AI civility toolkit** complements traditional moderation by providing opt-in, user-controlled support before posting (e.g., clearer framing) and filtering of incoming content. Relatedly, **ad-hominem detection** could provide immediate feedback and ranking signals that help keep discussion focused on arguments by reducing personal attacks. To protect crowd-driven systems from manipulation, **automated detection of filter algorithm manipulation** would combine credibility scoring, anomaly detection, reputation thresholds, and network analysis to identify artificial engagement. For courts and journalists, the **crowd media collage synthesizer** would combine overlapping eyewitness recordings into synchronized accounts while preserving attribution to each original source. The **selective-scope immutable stream ledger** would complement this by authenticating viral eyewitness videos only when risk signals trigger it, blocking the viral spread of tampered footage. Finally, **resolving popularity and majority bias** would give users the option to evaluate anonymized content independently of its creator's identity or popularity, helping new and underrepresented voices receive attention based on the perceived quality of their contributions. A **thread-privacy toggle** would complement these mechanisms by allowing users to control the privacy level of their post, making higher-risk conversations safer without requiring a private conversation.

This white paper's overall objective is to explore how we can deliberately design an ecosystem of signals, incentives, user controls, and integrity tools that make high-quality democratic discourse easier, more rewarding, and more resilient. The design of social media should flow from democratic values – including pluralism, free expression, equal participation, public justification, transparency, collective self-rule, and human authentication – and will require contributions from the humanities, the sciences, and engineering. We hope this white paper provides a starting point for a broader discussion.

# Table of Contents

# Introduction

Social media have revolutionized human communication (boyd 2010). Every one of us can now broadcast opinions and insights publicly in an instant. Depending on likes and shares, our messages can reach millions of people. The advent of social media was initially rightly seen as a liberating development. It promised more shared experiences, enabling people to better connect and understand each other. It also promised a more participatory political discourse nationally and internationally.

A classic concept of a space for democratic discourse is the “public sphere”: a space in which citizens can deliberate about matters of common concern outside direct state or market control (Habermas 1989). The internet frees people from the requirement of physical presence in a town square or salon and enables the creation of social media as much larger and more accessible digital public spaces for discourse. They can enable us to share our insights and help each other think, elevating our collective intelligence (Benkler 2006). Social media can also help citizens form social movements, and have in fact enabled movements to rapidly mobilize and gain visibility (though they also created new vulnerabilities to fragmentation and surveillance; Tufekci 2017). For all these reasons, the promise of social media ought to be very much alive.

Over the past decade, however, the perils of social media have come to the fore. Social media are increasingly seen as responsible for social isolation (rather than connection; Primack et al. 2017), political polarization (through filter bubbles and echo chambers; Cinelli et al. 2021, Sunstein 2017, Bakshy 2015, Haidt & Bail ongoing), mental health issues (Keles et al. 2020, Haidt 2024), misinformation (unreliable and misleading information; Lazer et al. 2018), and disinformation (organized messaging campaigns with manipulative intent; Bradshaw and Howard 2019). A systematic review of 496 articles on digital media and democracy reports a mixed picture: digital media were associated with greater political participation and information consumption in some contexts, but also with declining political trust, rising populism, and increasing polarization, especially in established democracies (Lorenz-Spreen et al. 2022).

Rather than connecting us in the right ways, the social media we currently have seem to isolate us as individuals, divide us into artificial tribes, polarize our political opinions, and sometimes bring out our worst impulses—making us lose our civility or enabling us to cause serious harm to other people’s livelihood and safety. Social-media posts expressing animosity toward political out-groups receive especially high engagement (Rathje et al. 2021). False news on Twitter has been shown to diffuse farther, faster, more deeply, and more broadly than true news, with especially strong effects for political news (Vosoughi et al. 2018).

Some feel that these negative outcomes are inevitable effects of heightened connectedness, and that social media simply hold a mirror to us, revealing ugly truths about human nature. However, there is no reason for such “Lord of the Flies” fatalism. The social dynamics on social media reflect not just human nature but importantly also the “game design” of current platforms. One theory suggests that the way current platforms are designed amplifies natural tendencies toward moral-emotional expression and in-group signaling (Brady et al. 2020).

This means that the architecture of existing social media must be re-evaluated (Landemore et al. 2021). Academic studies have focused on documenting antisocial behaviors—misinformation, hate speech, and outrage. More work is needed on how platforms can be designed to encourage prosocial behavior, for example, by facilitating a sense of community and by fostering collective problem-solving (Dörr et al. 2025). It is not at all clear that the particular social media platforms that dominate today show us the limits of what is possible.

Social media platforms such as Facebook, X, TikTok, Instagram, and Snapchat each started with an inspired initial design and have adjusted that design over the years to increase advertising revenue. This incentive structure gives rise to a business model that has been labeled "surveillance capitalism," in which platforms extract behavioral data, infer users' interests and intentions, and monetize through prediction and targeting (Zuboff 2019).

Companies are centrally motivated to increase "engagement" because it is the engine that drives ad revenue, which in turn sustains or grows the platform's valuation (Brady et al. 2026). Users produce and consume content. Advertisers pay for access to users' attention and data. To maximize ad revenue, platforms must maximize both the quantity of attention available (time on site, number of sessions) and the quality of targeting (how precisely ads can be matched to users' inferred interests and intentions). User engagement contributes to both. Engagement is also essential for democratic discourse, but must be defined in terms of meaningful exchanges of facts and opinions that provide substantive value to the individual and to the community: by improving the discourse. The industry's narrow focus on optimizing engagement for ad revenue is in tension with the goal of improving social media for democratic discourse.

In response to mounting criticism that optimizing engagement alone has negative consequences, such as amplifying outrage and causing addiction, some companies now experiment with alternative, more responsible metrics that attempt to capture "well-being" or give "meaningful interactions" more weight in overall engagement. Work on recommender-system alignment frames this as a question of optimization targets: platforms can, in principle, optimize not only for clicks or time spent, but also for values such as diversity, fairness, well-being, factual accuracy, and long-term user benefit (Stray et al. 2021). However, the metrics actually used remain constrained by narrow economic incentives: if a change to social media design reduces ad impressions too much, it is unlikely to persist or be further developed. Because the social media we have today are the product of ad-revenue maximization, their shortcomings tell us little about any inherent limitations of connecting people at a global scale.

New visions for the game design of social media (e.g., Weyl et al. 2025) are relevant not only to governments and non-profit organizations exploring non-commercial forms of social media, but also to social media companies seeking sustainable and prosocial business models. Platforms designed to foster prosocial behavior and democratic participation may generate more sustainable engagement over time, building long-term user trust and satisfaction. Designing social media for democratic discourse, then, could be a pathway to long-term commercial success for the industry.

Building social media for democratic discourse requires, first, an alternative vision and, second, an alternative funding or business model to ensure that design is driven by democratic values rather than ad revenue. Here, we focus on the first step: the alternative vision. The second step is beyond the scope of this white paper, but, briefly, alternative models include: (1) a *public-service or publicly funded model* (conceptualizing social media as part of a country's democratic infrastructure), (2) a *nonprofit, donor-supported model* (like Wikimedia or Mastodon), (3) a *subscription or membership model* (commercial, but users pay directly, so the platform is rewarded for long-run satisfaction, trust, useful moderation, reliability, and features people actually value), (4) a *cooperative or member-owned model* (a jointly owned and democratically controlled enterprise), and (5) a *protocol-and-service model* (where the social network is built as open infrastructure and monetization can move to the service layer).

Improving social media for democratic discourse is among the most important challenges of our time. Why has little progress been made? Part of the paralysis results from the fact that most of us feel that it's not our job to think about the design of social media. But if the tech companies have misaligned incentives and governments lack the resources and expertise to develop substantial alternatives, then who can make a difference? Developing a better vision seems daunting to many of us because it requires specialized expertise we do not have. For example, a political scientist may feel that they lack expertise in software engineering to get involved. A software engineer, similarly, may feel that they lack the philosophical, psychological, or sociological knowledge to envision improvements that would benefit democratic discourse. Designing social media for democratic discourse requires us to creatively connect many different kinds of expertise and ingenuity.

It is time that people from all fields and walks of life took on the creative challenge of envisioning social media the way they should be designed to serve democratic discourse. We need the ingenuity and technical savvy of people from the social media industry in this conversation. However, we also need people who are not subject to the incentive structure of social media companies, and who bring radically different perspectives and expertise to envision social media with a different set of goals. Creating a digital public sphere for democratic discourse will require a rethinking of social media from first principles. The humanities, the sciences, and engineering all have key roles to play.

We are a lab of cognitive computational neuroscientists, with expertise in psychology, cognitive science, computer science, and neuroscience. As academics and citizens, we want to contribute and help broaden the conversation. In this white paper, we develop ideas for mechanisms that could be implemented on social media to improve democratic discourse. Each proposed mechanism is described on a single page in the form of a table to make the ideas as accessible as possible and emphasize their modular nature. Each mechanism can be considered in isolation, and many of them make complementary contributions. Each mechanism description was contributed by one or several of the authors of this paper as indicated by the author list in the beginning of each section. Some of the ideas are novel; most of them are related to previous developments, as reflected in the references provided in each table. The paper serves both as an entry point into the literature on social media mechanisms and as a source of novel ideas.

We hope this set of proposed mechanisms provides a starting point for a broad discussion involving non-experts as well as experts, journalists as well as academics, and citizens from many walks of life. If you would like to help start a discussion or contribute your own ideas to future versions of this paper, please contact one of the corresponding authors (NK, TJ, ZL).

# Mechanisms

## Feed discovery & customization

*How a user discovers and actively curates content exposure*

Virtual travel through geographic location filtering - *Kedar Garzón Gupta*

Stance-based search - *JohnMark Taylor*

Communal feed curation - *Kedar Garzón Gupta*

Natural language feed algorithm tuning - *JohnMark Taylor, Patrick Stinson*

Open-source forkable recommenders - *Fan Cheng*

Daily algorithm ballot - *Amirhossein Farzmahdi*

Content-engagement summary and feed recalibration - *JohnMark Taylor, Zi Qi Liow*

Stream networks - *Nikolaus Kriegeskorte*

# Virtual travel through geographic location filtering

**Description**

The majority of U.S. adults regularly get news from social media platforms, through both direct links to news sites and more informal, user-generated discussions about current events. Platforms typically tune a user's feed to their geographic location, making clear the particular relevance of local events. However, users who want to view content from other regions of the world are often limited to very basic search functionality. With a clickable map, users could select fine-grained locations, such as specific towns or cities, and populate their feed with live content from that place. This virtual travel feature would enable people to instantly view current local news stories in a region of interest, along with the surrounding conversation. Users could explore events in another country or compare regional perspectives on a major nationwide story where they live. By navigating geographic space worldwide, users could sample a variety of cultural and political perspectives on locally occurring events.

**Promise**

- Virtual travel provides more direct access to the lived experience of people in other places
- Connects users to people across the world, including those in different cities, countries, continents, and cultures, as well as in regions affected by natural catastrophes and wars
- Clickable map for intuitive visual navigation
- Improved sampling of local, cultural, and political perspectives

**Challenges**

Social media content is rarely geotagged—some estimates place Twitter at 3% and Instagram at 25%. However, machine learning-driven approaches can extract location from text and image content for 50-90% of untagged items. In addition, local events can be detected as spikes in the overall rate of geotagging. Users' locations can also be inferred from those of their closest friends or followers

- Geotagging is more common in urban areas, so localizing content to rural areas is more difficult
- Location can be spoofed with a VPN, so platforms may need to validate with inferential methods
- Users may geotag posts while they are traveling, so "localness" must be attributed carefully

**Related ideas** Cross-lingual bridge layer

**References** Pew Research Center 2024, Flatow et al. 2015, Han et al. 2014, Harrigian 2018, Hecht and Stephens 2014, Krumm and Horvitz 2015, Rout et al. 2013, Johnson et al. 2016

## Stance-based search

### Description

Current search algorithms are generally based on the presence of specific words rather than the stance being advocated. This makes it challenging to find posts advocating for specific positions, often requiring guesswork about which words to include in a search query. A search algorithm sensitive to stance content could make it easier to locate arguments for and against specific views. Additional functionality could automatically curate a range of positive, negative, or neutral opinions on a given stance, as desired.

### Promise

- Facilitate finding arguments for and against certain positions
- Method for users to escape echo chambers

### Challenges

- May require novel search algorithms (search could incorporate a post's embedding in an LLM)
- Some users might not want their posts to be available in this way. This could be addressed by allowing opt-out in user settings

**Related ideas** Devil's advocate button, live opinion mapping for current events, number-free feed, Cognicons, resolving popularity and majority bias

**References** Bozdag and van den Hoven 2015, Sharma et al. 2024, Xiong et al. 2024

## Communal feed curation

**Description**

Allow a community of users to curate a feed or algorithm based on their shared value system. This differs from existing "groups" functionality on platforms such as Facebook, where content must be authored by group members. A community can be defined as loosely or as tightly as possible—it may revolve around a niche shared interest or a broader sense of intellectual cohesion. Communities could choose to publicly release their curated feeds, accompanied by a tagline or mission statement, allowing other users to subscribe. By actively shaping the content that proliferates across their social networks and explicitly articulating community values, users become more aware of the content they encounter and less susceptible to unknowingly falling into echo chambers.

**Promise**

- Community-based feed curation and algorithm tuning
- Outside users can subscribe to this communal feed

**Challenges**

- Existing feedback and reaction mechanisms may not sufficiently redirect the algorithm towards community values. This could be addressed using *Natural Language Feed Algorithm Tuning*

**Related ideas** Natural language feed algorithm tuning, daily algorithm ballot

**References** He et al. 2023, Burton et al. 2024

## Natural language feed algorithm tuning

**Description**

Users would have the ability to provide natural language instructions to tune their feed algorithm through an LLM interface (e.g., “show me more posts about PEPFAR”). A user’s settings would include a log of these instructions, which could be retroactively modified if the user changes their mind. The log would also feature “up” and “down” controls next to each instruction, allowing users to adjust the magnitude of changes if the algorithm overcorrects or undercorrects. This capability is partly motivated by a common failure mode in which users habitually engage with low-quality but “addictive” content, leading the algorithm to increase the prominence of such content in the feed. Users could instead downregulate this content in line with their conscious goals for social media consumption.

**Promise**

- Gives the user more transparent and flexible control over the filtering algorithm using an intuitive interface
- Control over the algorithm is based on the user’s conscious preferences and values, rather than habitual behavior
- No need for hard-coded “dials” for tuning the filter; the user has total flexibility

**Challenges**

- Designing reliable LLM-tunable filtering algorithms could be challenging

**Related ideas** Communal feed curation, daily algorithm ballot (DAB), open-source forkable recommenders

**References** Bhargava et al. 2019, König 2024, Burrell et al. 2019

# Open-source forkable recommenders

## Description

Let users choose, tune, or "fork" the ranking algorithms that build their feeds. The default recommender still exists, but people can (a) install alternative open-source recommenders from an audited marketplace (Yang et al. 2018; Salah et al. 2020) and (b) view counterfactual feeds generated by other algorithms (Tan et al. 2021; Hosseinmardi et al. 2024), so they can immediately see how different objectives change what they read (Zheng and Wang 2022). This mechanism unites three layers that current platforms expose only piecemeal: genuine algorithm choice, interactive counterfactual views, and transparent, user-friendly tuning with side-by-side differences.

```
Feed Mode:  [ Default (Engagement ON) ]  |  [ Counterfactual: No-Engagement ]
View:       [ Side-by-Side Diff ✓ ]  [ Single Feed ]

Default Rank  |  Counterfactual Rank   Post Card (hover for details)

 1  Post A    |   4  Post A   ↓↓↓       [Post A]
               \                         2h ago • 1.2k likes • 120 comments
 2  Post B    |   2  Post B   –          ⓘ Why this post?
               \                            • High engagement in your network (+0.32)
 3  Post C    |   7  Post C   ↓↓↓         • Similar to topics you clicked last week (+0.21)
               \                            • Recent (2h) (+0.12)
 4  Post D    |   1  Post D   ↑↑↑       [ Show alternative ranking breakdown ▸ ]
               \
 5  Post E    |   3  Post E   ↑          ▼ Alternative Ranking (No-Engagement)
               \                            Rank if engagement=0: #4  (Δ -3)
```

## Promise

- Turns "algorithmic transparency" into user control
- Competing algorithms spark innovation and independent audits
- Counterfactual feeds build literacy about amplification dynamics
- Open ecosystem reduces single-platform dependence and lock-in

## Challenges

- Interface overload could confuse users (but could provide preset bundles, simple mode toggle)
- Third-party algorithms may leak or misuse data (which could be addressed using sandboxed APIs, mandatory audits)
- Few users switch without clear benefits (which might be alleviated by offering preset options, social proof)

**Related ideas** Natural language feed algorithm tuning, automated detection of filter algorithm manipulation, daily algorithm ballot (DAB), communal feed curation

**References** Yang et al. 2018, Salah et al. 2020, Tan et al. 2021, Zheng and Wang 2022, Hosseinmardi et al. 2024

# Daily algorithm ballot

**Description**

Imagine opening your social app at breakfast and seeing a tiny prompt, a one-question ballot asking which topics you would like to prioritize in today's feed. You tap "local news" over "viral posts" and close it. Thousands of people in your region (with an adjustable reach limit) do the same. By midday, a public banner appears: Today's vote: +30% local news, +15% cross-partisan viewpoints, –20% viral posts. From then until the next morning, everyone's feed leans slightly toward these shared priorities. The duration of the effect can also be adjusted. The daily algorithm ballot invites users to take part in a daily, optional poll. It asks simple questions about feed preferences: chronological order, ideological diversity, or calmer conversation. The platform aggregates responses by time zone and tunes a few settings in the ranking system for the next 24 hours. Your individual follows and interests still guide your feed, but now the community's voice helps shape the overall balance. The daily algorithm ballot turns feed curation from a hidden, engagement-optimized formula into a more transparent one.

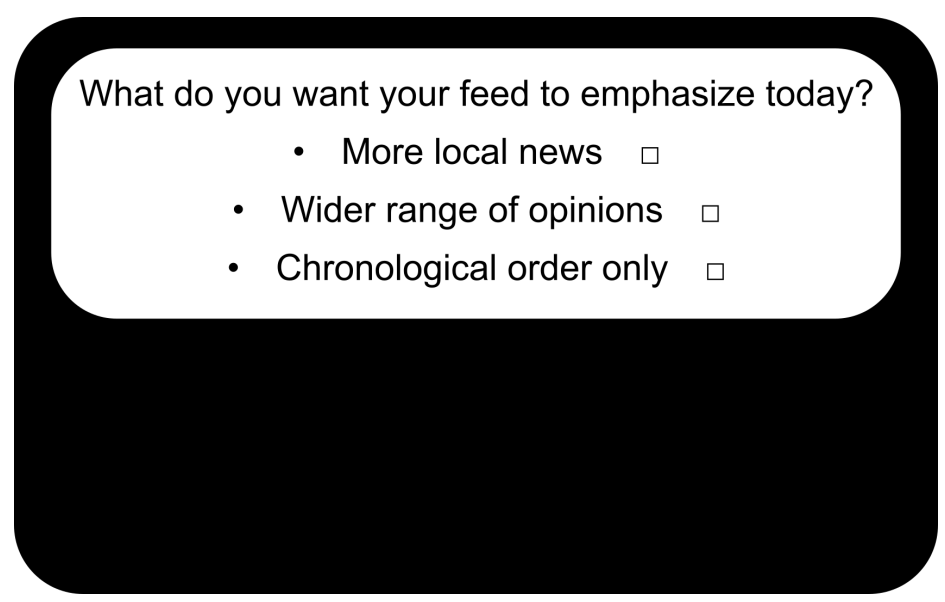

**Promise**

- Community engagement
- Transparency
- Diverse perspectives
- Reduced viral pressure
- User empowerment

**Challenges**

- Low participation (embed the one-click ballot in the feed header)
- Vote manipulation (cap ballots using verified accounts)
- Engagement decline (keep a baseline of personalized content)
- Complexity management (limit to plain language)

**Related ideas** Natural language feed algorithm tuning, communal feed curation, collaborative fact-checking

**References** Helberger 2021, Lorenz-Spreen et al. 2020, Törnberg et al. 2023, Lasser and Poechhacker 2025

## Content-engagement summary and feed recalibration

### Description

Users can access a summary of the content they have viewed, posted, or engaged with over a given time period. This summary would include descriptive metrics of political or personal relevance, such as ideological diversity (Munson et al. 2013) or time spent "doom-scrolling." The summary could additionally include "word cloud" visualizations of topics that were most prominent in the user's engagement habits. Borrowing the visualization technique from Nagulendra et al. (2014), this word cloud could also include topics that were prominent for others, but were not included in the user's experience, using a bubble to divide the two (making the operation of the "filter bubble" more salient).

This summary could additionally include descriptions of how the algorithm categorizes the user, allowing the user to correct any "misconceptions" if desired. Users could ask specific questions about their engagement habits, such as "why have I been seeing more posts about X lately" or "how has my ideological diversity score changed over the past month?" making the feature actionable rather than merely descriptive. This summary could be both accessible on demand, and provided automatically in a periodic review (inspired by the popular "Spotify Wrapped" feature that summarizes music listening habits).

Finally, users could specify goals for themselves (e.g., consume more politically diverse content). The discrepancy between the summary of actual engagement and the user-specified goals could form the basis for an automatic recalibration of the feed. Users could also share or compare their goals with others, turning this into a collective accountability mechanism.

### Promise

- Gives user access to actionable information for adjusting their social media habits if desired
- Increased awareness of the "filter bubble" and how it is constructed
- Design of summary itself could give implicit "nudges" (e.g., that the content they consume is dominated by one political perspective, or that they spend more time on the app than they would like)

### Challenges

- Ideological diversity is difficult to quantify, and requires classifying content along political dimensions
- Distinguishing between doom-scrolling and engaged, productive scrolling

**Related ideas** Natural language feed algorithm tuning, open-source forkable recommenders

**References** Munson et al. 2013, Nagulendra et al. 2014, Xing et al. 2014

## Stream networks

**Description**

A stream network is a network of connected feeds (called streams). Each stream is controlled by an individual user. A user can create a stream to capture a particular kind of content (e.g., news, developments in a particular area of science or technology, personal news within the user's circle of friends). One user can set up and control multiple streams. A stream is a mixture of other streams, which provide the incoming content. Each post in a stream has a priority rating. The priority rating of a post is a function of the priority ratings of the post in all incoming streams, e.g., a weighted average. A stream selects posts from all incoming streams, based on their combined priority ratings, their recency, their content, the rate of the stream (posts per minute), and rate limitations imposed on the incoming streams. Streams can be explicitly defined either using a domain-specific language or by visual programming and can employ machine learning and artificial intelligence elements, which enable them to learn to make better decisions on posts for their owner.

A stream network is a distributed computer that processes cultural items created by people and enables a community to collaboratively create a culture by curating the flow of content. The network of streams provides a decentralized mechanism for democratic discourse, where no single platform algorithm controls public attention, either at the level of evaluative signals or feed algorithms. The architecture extends proposals for social-media middleware, protocol-based decentralization, and user-selectable feed algorithms by making curation itself a recursively composable, user-owned social activity rather than a service provided by either a platform or a separate algorithm vendor. A stream network constitutes a form of collective intelligence and networked public sphere in which both human editorial judgments and computer-automated algorithmic selection are distributed across users and their hardware and computationally recombined, rather than concentrated in a single platform's ranking system.

**Promise**

- Highly general conception that provides a superset of existing social-web mechanisms
- Each user has complete control over their streams and may share them with the community
- Multiple streams give user control over time spent on different interests

**Challenges**

- Need to develop domain-specific language and visual programming environment for streams
- Need to provide more accessible forms of stream design, e.g., simple selection of incoming streams and binary or continuous user evaluation of posts for machine learning of the post-selection function
- Need to achieve interoperability of streams and the traditional social web

**Related ideas** Open-source forkable recommenders

**References** Altay et al. 2025, Balabanović and Shoham 1997, Goldberg et al. 1992, Fukuyama et al. 2021, Hogg et al. 2024, Kahn 1974, Kleppmann et al. 2024, Kriegeskorte 2012, Liu et al. 2025, Malone et al. 2010, Resnick et al. 1994

# Content contextualization & understanding

*Features that help the user process information and avoid echo chambers*

Cross-lingual bridge layer - *Fan Cheng*

Live opinion mapping for current events - *Kedar Garzón Gupta*

Conversation graph for meaningful discourse - *Amirhossein Farzmahdi*

Timeline generation for controversial societal issues - *Pinyuan Feng*

MatchAcross: connecting users across ideological divides via social proximity - *Akihito Maruya*

Interactive bubble-chart opinion clustering - *Zi Qi Liow*

## Cross-lingual bridge layer

**Description**

Unlike click-to-translate buttons that most users ignore, the platform could offer an opt-in layer that auto-translates any post or comment outside of a user's fluent languages. Original and translated text can appear either inline or in a split-pane view, letting mixed-language feeds stay readable even at high scroll speeds. Users set granular rules, e.g., "never translate English or Japanese; translate everything else into Spanish." Translation quality carries model-confidence badges and crowd-suggested fixes; readers can propose better wording, add cultural notes, or pin terms to a shared mini-glossary. For live streams or rapid threads, low-latency machine translation plus crowd/LLM smoothing generates rolling subtitles. By showing people other populations' words directly—without forcing authors to switch languages—this layer makes cross-language reading habitual and reduces misunderstandings.

**Promise**

- One-click rules eliminate repetitive "Translate" clicks, increasing cross-language reading
- Side-by-side originals preserve nuance and accountability
- User rules reduce clutter, respect bilingual/multilingual habits
- Community edits improve machine translation quality and cultural context over time
- Bridges diaspora and global conversations, enriching deliberation

**Challenges**

- Machine translation errors distort meaning. (This could be addressed by crowd fixes and confidence labels)
- Slang/dialect gaps (addressable by community glossaries and domain tuning)
- Added interface complexity (addressable by smart defaults and hover reveals)
- Abuse via mistranslation trolling (addressable by version history and moderation)
- Latency in live contexts (addressable by progressive refinement and caching)

**Related ideas** AI civility toolkit, AI fact-checking

**References** Mendelsohn et al. 2023

## Live opinion mapping for current events

**Description**

For a developing news story, an opinion map of users' posts could be automatically generated. This would allow users to interact with a story and view/sort others' posts according to desired filters—e.g., political leaning, controversiality, geographic location, humor/memes, or modality (text, image, etc.). This way, the conversation surrounding a current event is centralized and navigable to allow users to sample a variety of perspectives by their metrics of choice.

**Promise**

- Live, interactive map of people's opinions
- Diversity of filters for sorting posts, e.g., by political content, trendiness, location, modality

**Challenges**

- Classifying potentially millions of posts in real time is technically challenging. However, this might be addressed with ML-based clustering algorithms (e.g. Knittel et al. 2022)

**Related ideas** Conversation graph for meaningful discourse

**References** Knittel et al. 2022, Faridani et al. 2010, Cortis and Davis 2021, Wu et al. 2014, Garimella et al. 2015

## Conversation graph for meaningful discourse

**Description**

Structured conversations transform social media from a focus on viral attention toward fostering mutual understanding and healthier democratic exchanges. Rather than confusing comment threads, discussions are organized visually like a graph. Imagine the original post as a point in the graph, anchoring the main topic. Each reply forms a branch, explicitly labeled (either by users or AI-suggested) as a supporting point, a counterargument, or a clarifying question. Further replies create additional branches, making it easy to visually track and understand the conversation's evolution and the relationships between ideas. This structure encourages thoughtful, coherent dialogue and clearly presents diverse viewpoints, helping users navigate complex topics more effectively. Platforms such as Kialo implement similar graph-based argument mapping to enhance classroom debate and critical thinking, demonstrating the effectiveness of structured conversations in promoting meaningful discussion.

**Promise**

- Graph views reveal how points, counters, and questions connect, making debates easier to follow
- Separate branches ensure supportive, opposing, and inquisitive replies all stay visible
- Healthier tone and better learning outcomes

**Challenges**

- The extra effort to tag replies which can be reduced with AI-suggested labels
- Deliberate mislabelling
- Huge graphs get messy. This could be addressed by auto-collapsing low-activity branches

**Related ideas** Discourse score, resolving popularity and majority bias, milestone badges for high-quality participation

**References** Ouyang et al. 2024, Anastasiou and Liddo 2025, Irani et al. 2025

## Timeline generation for controversial societal issues

**Description**

Important social and political debates often get lost in a swirl of outrage and fleeting viral moments, making it difficult for users to understand the full context or history of an issue. This concern could be addressed by automatically assembling clear, chronological summaries of key events, policy changes, viral posts, and turning points for major public debates. Instead of only seeing the latest heated comment or news flash, users can scroll through a structured timeline that highlights how discussions, opinions, and facts have evolved over time, complete with links to original sources and concise summaries of multiple viewpoints. This approach helps users catch up quickly, make sense of complex issues, and engage in conversations with a more informed, big-picture perspective.

**Promise**

- Reduces misinformation
- Improves civic literacy

**Challenges**

- Social issues often evolve rapidly, with new facts or interpretations emerging. We can mark timeline events as “under review” or “updated” when facts are still emerging. Also, an integrative verification system should be developed, such as crowdsourcing review and flagging of disputed entries
- Summarizing complex events risks oversimplification or factual mistakes, especially in heated debates. It requires every timeline entry to include verifiable citations

**Related ideas** Crowd media collage synthesizer, selective-scope immutable stream ledger, AI fact-checking, live opinion mapping for current events

**References** Vosoughi et al. 2018, Porter and Wood 2021, Pennycook and Rand 2019, Knittel et al. 2022, Wu et al. 2014

## MatchAcross: connecting users across ideological divides via social proximity

**Description**

Social media often increases political polarization and social isolation by showing users mostly opinions they already agree with. Research shows that simply exposing people to opposing views—especially from strangers—can sometimes backfire and further entrench division (Bail et al. 2018). However, a meta-analysis of 115 studies found that awareness that in-group members have friends in another group—a phenomenon known as extended contact—reliably improves attitudes toward out-group members (Zhou et al. 2019).

Inspired by this finding, we propose an AI-based system designed to help users engage with a broader range of perspectives in a constructive and empathetic way. The system encourages conversations with people who hold different views, either within a user's existing network or just outside of it—such as a friend-of-a-friend—so that users know there is a social connection. It uses data such as liked posts, viewed content, and friend networks to estimate each user's position and presents this on a simple, intuitive visual map—similar in spirit to the Echo Chamber Map developed by the Polarization Lab at Duke University. Such a map suggests guided interactions that foster thoughtful dialogue and reduce defensiveness. Users also remain in control of how far across the ideological spectrum they wish to engage, creating a flexible and personalized experience.

**Promise**

- By exposing diverse viewpoints, the system encourages empathy and understanding rather than conflict
- Encourages cross-ideological interactions
- Visualization of one's position with respect to others may help users reflect on their views, biases, and how they compare to others

**Challenges**

- Mapping people's opinions based on a few sources (e.g., likes and friends list) might be difficult
- Oversimplifying or mislabeling users, leading to mistrust or incorrect matches
- Users might feel labeled or ranked

**Related ideas** Live opinion mapping for current events, stance-based search, conversation graph for meaningful discourse, virtual travel through geographic location filtering

**References** Bail et al. 2018, Zhou et al. 2019

## Interactive bubble-chart opinion clustering

**Description**

A real-time interactive visualization that maps individual and collective stances on a specific issue. The bubble chart is an opt-in decision added by the creator to their post, not something that appears on every post. Each user can locate themselves ideologically by manually placing their "dot" on the chart, with axes representing different dimensions of the issue. Placement on an axis represents where the user stands personally. Similar dots form clusters, visually revealing the diversity or homogeneity of opinion. Users can click on any cluster to: (A) view profiles, (B) enter one-on-one messaging or group chats with users on the chart to understand differing or shared views, and (C) update their "dot" position post-conversation/over time to reflect changed understanding or reinforced beliefs.

The axes are chosen by the post creator. Alternatively, the AI can suggest axes based on the post's content, using NLP to detect the underlying tension(s) the post raises. The AI matches the post's core claim and entities against a bank of known opposing-value pairs via embedding similarity. Creators can edit the axis dimensions to better fit their content – some may use one axis, while others may be multi-dimensional.

**Example:** For a post about whether more funding should be distributed to the police department or to social services, the following bubble chart may be displayed.

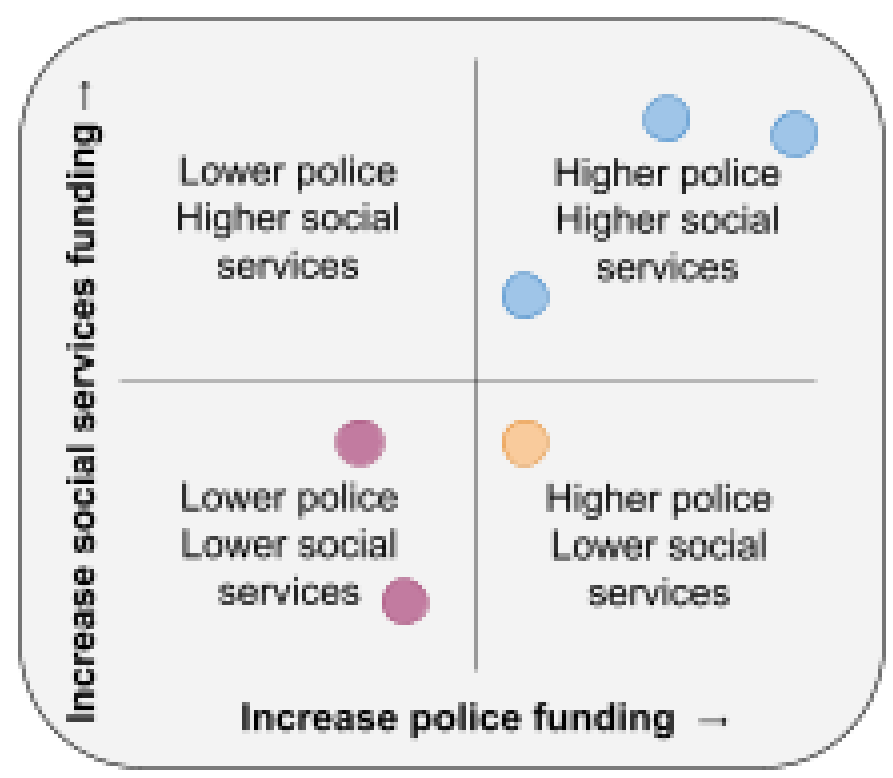

**Promise**

- Complement binary reactions (e.g., like/dislike; thumbs up/down) with nuanced positioning
- Make echo chambers and outliers visually perceptible
- Foster middle ground conversations about complex topics

**Challenges**

- Privacy concerns (Solution: Post your dot publicly, privately, or with friends; accept or decline chat requests)
- Posts with low engagement may not have enough dots to reveal meaningful clusters (Solution: Delay full chart visibility until either N responses or T time has passed)

**Related ideas** MatchAcross, live opinion mapping for current events

**References** Bakshy, Messing, and Adamic 2015

## Engaging & reacting

*How the user interacts with content while minimizing bias*

Cognicons - *Nikolaus Kriegeskorte, Trenton Jerde*

Probability ratings - *Nikolaus Kriegeskorte, Trenton Jerde, Patrick Stinson*

Devil’s advocate button - *JohnMark Taylor*

Number-free feed - *Amirhossein Farzmahdi*

Thread-privacy toggle - *JohnMark Taylor*

# Cognicons

**Description**

Emoticons offer an efficient (one-click) mechanism for sharing emotional responses to social media posts. Capturing emotional responses enables feed algorithms to amplify emotionally engaging content. In order to amplify posts that constitute good contributions to rational discourse (where the goal is to understand the world together) and/or to democratic discourse (where the goal is to negotiate a shared vision for the future), it would be useful to capture cognitive responses through a similar one-click mechanism. For example, users should be able to share a judgment that a post (or the content it points to) is correct or incorrect, or that it provides an interesting perspective that we don't necessarily agree with. It should be easy (1) to flag false information, (2) to share high-quality content without endorsing it, and (3) to reward our friends for sharing things we don't agree with, such as opinions from political opponents, when they are thought-provoking and carefully argued. Emoticons emotionalize the sharing culture and may contribute to political polarization and the formation of echo chambers. Cognicons are icons similar to emoticons that enable us to share our cognitive responses (see figure below). LinkedIn's user-selected "Insightful" and "Curious" reactions are a step in this direction. Importantly, cognicons should not be labels of posts chosen by an algorithm or opaque group of moderators. Instead, like emoticons, cognicons should be available to users for annotating posts.

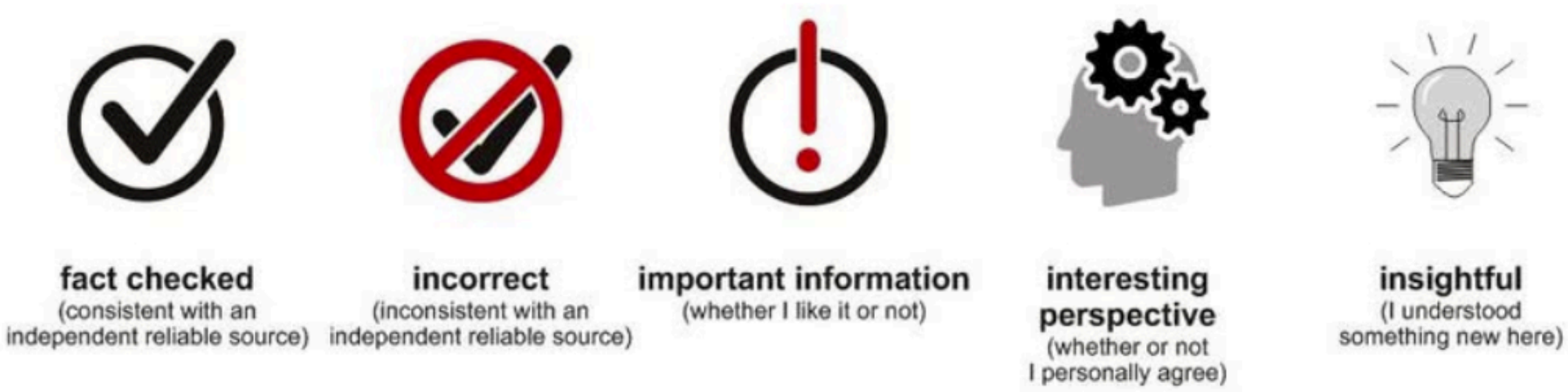

**Promise**

- Efficient sharing of cognitive responses
- Can help us think well together and explore a diversity of viewpoints
- Important counterweight to emoticons which emotionalize the sharing culture

**Challenges**

- A large number of cognicons might be confusing and difficult to understand (choose a reasonable number to balance nuance and simplicity)

**Related ideas** Probability ratings

**References** Avram et al. 2020, Martel et al. 2020, Pennycook et al. 2021, Jahanbakhsh et al. 2021, Pennycook and Rand 2022, Stoeckel et al. 2024

# Probability ratings

**Description**

Social media provide efficient mechanisms for information sharing in large communities. A key challenge is to enable the community to evaluate the reliability of the information. Centralized fact-checking faces challenges of legitimacy and scale whether implemented using human judgment (i.e. moderators or fact checkers) and/or AI. If users are given the option to annotate posts with one-click probability ratings on a continuous scale from 0 to 1, they can efficiently share their judgments of the probability that a given post is factually correct. Other users can be presented with the distribution of ratings a post received. The distribution of ratings from trusted associates can be shown separately from the full distribution. The rating distribution could be corrected for representativeness of the distribution of political viewpoints. Aggregates of the ratings can also be presented, such as the mean, the logit mean, or a probabilistic collective inference of the probability of a claim, leveraging probabilistic models of the rating behavior of individual users to account for user accuracy and calibration.

**Promise**

- Continuous ratings enable nuanced communication of judgments about the factual correctness of posts
- Collective inference algorithms promise accurate crowd-sourced inferences about the truth of claims
- More legitimate according to democratic principles than centralized fact-checking

**Challenges**

- Individual probability ratings are known to be overconfident on average (calibration can be estimated and corrected)
- Collective inference requires user models to capture miscalibration (efficient algorithms that scale to large sparse data exist, e.g. Stinson et al. 2025)
- Dependency among user judgments is difficult to model unless users have co-judged large volumes of posts (supervised calibration with truth labels can help calibrate collective inference)

**Related ideas** Discourse score, number-free feed, milestone badges for high-quality participation

**References** Genest and Zidek 1986, Ranjan and Gneiting 2010, Baron et al. 2014, Tetlock et al. 2014, Mellers et al. 2014, Satopää et al. 2014, Satopää et al. 2016, Satopää et al. 2017, Tschiatschek et al. 2018, Roitero et al. 2020, Allen et al. 2021, Saeed et al. 2022, Panizza et al. 2023, Stein et al. 2023, Barbera et al. 2024, Drolsbach et al. 2024, Stinson et al. 2025

## Devil’s advocate button

**Description**

All posts have a “devil’s advocate button” that provides alternative viewpoints to the opinion expressed by the post. These can take the form of (1) pulling up example posts or articles articulating an alternative view (algorithmically selected based on relevance and quality), (2) giving an AI-generated response articulating an alternative view, or (3) notifying some number of users who have voluntarily “registered” to be advocates of a position to come and argue their point. For (3), these users could be subject to a rating system vouching that they are respectful interlocutors, and they in turn would be able to rate the user who “summoned” them; disrespectful users would have this feature limited.

**Promise**

- Simple one-press “escape valve” from an echo chamber
- Educational aid for users who wish to learn the pros and cons of a position

**Challenges**

- Finding and selecting alternative views may be computationally challenging

**Related ideas** Thread privacy toggle, virtual travel through geographic location filtering, live opinion mapping for current events, communal feed curation, daily algorithm ballot (DAB)

**References** Bozdag and Van Den Hoven 2015, Pariser 2011, Nguyen et al. 2014, Resnick et al. 2013

## Number-free feed

**Description**

Imagine scrolling through your feed with no visible “like” counts, share numbers, or follower totals – just the content itself. You can still tap a “reaction” button after reading, but your response is recorded privately. It is feedback for the creator, not a public leaderboard. Behind the scenes, every reaction helps refine what you see next, but the emphasis is on engaging with ideas rather than chasing popularity. Without visible counts, creators focus on substance: well-crafted arguments, insightful commentary, or creative storytelling. Readers aren’t biased by an invisible “crowdthink” effect, and newcomers aren’t discouraged by zero counts. The building blocks already exist. Instagram experimented for two years with hidden like totals, and now anyone can opt to hide them. X has proposed muting engagement counts to simplify timelines. Both moves acknowledge that reducing public metrics can improve the quality of interactions. In this environment, every post stands on its own merits, and every reaction becomes a private signal of genuine engagement.

**Promise**

- No public counts means fewer comparisons and healthier mood
- Posts compete on ideas, not on eye-catching counts
- Fresh voices aren’t dismissed for having “zero likes”
- Creators focus on substance instead of clickbait tactics
- Bots gain little by inflating invisible numbers

**Challenges**

- Harder to spot good content (but one could add user-given tags like “informative” to guide discovery)
- Creators lose proof for sponsors (mitigated by showing analytics dashboard)
- Opaque ranking if counts still drive the algorithm
- Could cut motivation: without visible likes, people might post less

**Related ideas** Cognicons, probability ratings, discourse score, resolving popularity and majority bias, milestone badges for high-quality participation

**References** Burnell et al. 2025, Lee and Kwon 2025, Wallace and Buil 2021

# Thread-privacy toggle

**Description**

The user is given options for setting the privacy level of a given post or thread, such as "any user", "followers only", "contacts of contacts", or "only participants in the conversation." This empowers users to control and guard against the risk of pile-ons from outside observers. It also gives users more fine-grained control than existing options (e.g., existing apps have a blanket option of "going private," which hides all of a user's posts from everyone except their followers). It also provides a way to voice and discuss "risky" opinions (e.g., that an employer or government might find objectionable) without creating an anonymous account.

**Promise**

- Providing post-level control over audience allows the user to escape "performative dialogue" when they desire
- Preventative measure against potentially undesirable "pile-ons"
- Provides a way to discuss controversial topics without the risk of punishment from employers or governments

**Challenges**

- Arguably redundant with DMs, which also provide privacy (but private posts can serve as a middle ground)
- Bad-faith actors could still screenshot and share responses (private posts should have a clearly visible indicator)

**Related ideas** Devil's advocate button, virtual travel through geographic location filtering, live opinion mapping for current events, communal feed curation, daily algorithm ballot (DAB)

**References** Misra et al. 2016, Trepte et al. 2021, Paramarta et al. 2018, Mondal et al. 2019

# Accuracy & civility

## *Tools for collaborative filtering and maintaining ecosystem health*

Game-aware design for democratic discourse - *Nikolaus Kriegeskorte, Trenton Jerde, Patrick Stinson*

Collaborative fact-checking - *Pinyuan Feng*

AI fact-checking - *Akihito Maruya*

Ad-hominem detection - *Nikolaus Kriegeskorte*

Crowd media collage synthesizer - *Fan Cheng*

Resolving popularity and majority bias - *Akihito Maruya*

Selective-scope immutable stream ledger - *Fan Cheng*

Automated detection of filter algorithm manipulation - *Amirhossein Farzmahdi*

Population-representative feeds - *Nikolaus Kriegeskorte*

Graded control of thread privacy and post visibility - *JohnMark Taylor & Nikolaus Kriegeskorte*

## Game-aware design for democratic discourse

**Description**

However a social network is designed, the resulting social dynamics can be understood as a game. Applying this lens requires analysis of the psychological dynamics of participation, including the rewards experienced by users. Social media participation is thought to be driven by people's need for social validation and status. When the design of the system is guided by the objective to increase ad revenue by increasing engagement, the game that results may be one where people strive to exhibit conformity with their in-group and prominent distinction from their out-group to raise their social status. This dynamic limits free expression and collective intelligence and may drive polarization (Bail 2021). Social validation and status are basic human needs that cannot be changed. However, what behaviors yield social validation and status depend entirely on the game set up by the social media design. For example, an alternative design in which ad-hominem attacks are flagged and posts that contain them have low visibility will effectively limit this particular undesirable form of exchange. Similarly, a design in which users use *cognicons* (alongside or instead of emoticons) to share responses to posts may emphasize rational and cognitive over emotional responses. Mechanisms like *cognicons* and *probability ratings* promise to set up a game in which social validation and status derive from thoughtful and original contributions to discourse. Probability ratings, in particular, could form the basis for a "game of truth" where users are validated and gain status to the extent that their probability judgments are accurate and well calibrated. More generally, social media design can be construed as the problem of how to set up a game that serves democratic desiderata, which include democratic participation, rational discourse, group deliberation, and collective intelligence.

**Promise**

- Center the individual motivations and needs and the resulting social dynamics
- Consider the creative challenge of designing a game that serves democratic discourse

**Challenges**

- Avoid heavy-handed social engineering (e.g., by scoring people and/or their contributions according to narrow criteria) that conflicts with the ideal of a versatile and general venue for many different kinds of exchange among free people
- Consider the multiple democratic desiderata (free expression, deliberation and collective intelligence, individual rights, participation), which may synergize in certain respects and pose tradeoffs in other respects (Bozdag and van den Hoven 2015)

**Related ideas** Cognicons, probability ratings, discourse score, number-free feed, conversation graph for meaningful discourse

**References** Tetlock et al. 2014, Bozdag and van den Hoven 2015, Mellers et al. 2015, Kleinberg and Raghavan 2019, Avram et al. 2020, Miller et al. 2020, Bail 2021, Brady et al. 2021, Stray 2021, Bak-Coleman et al. 2021, Pennycook and Rand 2022, Lorenz-Spreen et al. 2022, Sundar et al. 2024, Stinson et al. 2025, Milli et al. 2025, Dörr et al. 2025, Brady et al. 2026

## Collaborative fact-checking

**Description**

Today, fact-checking on social media often struggles with scale, partisanship, and lack of engagement. Platforms rely on a few methods, such as fact-checkers or opaque algorithms, while most users either ignore or distrust corrections. We can crowdsource the verification process to the user community in a structured, transparent way. Here, users work together to investigate claims, add context, cite sources, and vote on the accuracy of content. This approach not only expands fact-checking capacity, but also increases public trust and media literacy by involving diverse perspectives, making the process visible and open for scrutiny. Finally, collaborative fact-checking can help slow the spread of misinformation and foster a culture of shared responsibility for truth online.

**Promise**

- User engagement
- Transparency
- Rapid response

**Challenges**

- Lack of regular audits or monitoring may allow low-quality contributions to be rewarded. It requires multiple independent users to review and agree on fact-checks before they're marked as "verified"
- Expertise gap. We can address this by granting special status to users with proven expertise (e.g., scientists, journalists), making their input visible but not absolute

**Related ideas**

AI fact-checking, probability ratings, cognicons, selective-scope immutable stream ledger

**References**

Pennycook and Rand 2019

## AI fact-checking

**Description**

False information spreads more rapidly and widely than accurate content on social media, undermining the quality of public discourse (Vosoughi et al. 2018). Fact-checking can help curb misinformation (Porter and Wood 2021), but its reach is often limited. An AI-driven system could embed credible, contextual references directly into online content by identifying claims within posts, comments, and shared media, and linking them to relevant academic papers, fact-checking databases, or expert-reviewed articles. Users could select AI models reflecting different ideological perspectives, promoting transparency in source selection and exposure to diverse viewpoints. Such a system would support evidence-based engagement, enable users to assess credibility in real time, and contribute to a more informed and balanced digital public sphere.

**Promise**

- Real-time credibility assessment
- Promotes critical thinking
- Reduces polarization

**Challenges**

- Determining which source is credible in the end relies on the person
- Fact-checking is time-consuming

**Related ideas** Collaborative fact-checking

**References** Vosoughi et al. 2018, Porter and Wood 2021

## Ad-hominem detection

**Description**

Ad-hominem (“at the person”) describes contributions that judge other participants rather than the content being discussed. Mechanisms that enable flagging of contributions that veer into ad-hominem could help focus discourse on content. Ad-hominem contributions are considered undesirable for democratic discourse because they distract from the substance of the discourse and can intimidate participants, limiting their free expression and excluding potentially important insights. Ad-hominem flagging could be based on human judgments (e.g., there could be an ad-hominem *cognicon* for annotating posts) and/or on AI analysis of the content of posts. The flag could be binary or continuous, with the latter providing a nuanced signal that may help prevent derailing of discourse early when a binary flag would not yet be justified. The resulting evaluations of posts could be displayed with the posts to help participants stay clear of ad-hominem and to help other users assess the quality of the discourse. The ad-hominem evaluations could also be used for user-controlled ranking of content to be presented in feeds, enabling users to avoid discourse dominated by ad-hominem judgments.

**Promise**

- Can be used for immediate feedback to users to help sensitize them and save discourse by nipping ad-hominem in the bud
- Can be used to enable an audience to assess the nature of the discourse quickly
- Can be used in selecting posts for amplification, ensuring that widely visible discourse is not dominated by ad hominem

**Challenges**

- Requires human and/or AI judgment of what counts as ad-hominem
- Human moderator judgments of ad-hominem lack democratic legitimation and do not scale
- Human user judgments provide a legitimate crowd-sourced alternative, but could be weaponized by bad actors
- AI could contribute to ad-hominem detection but would require careful monitoring and minimization of bias

**Related ideas** Cognicons, AI civility toolkit

**References** Hosseini et al. 2017, Habernal et al. 2018, Delobelle et al. 2019, Patel et al. 2022, Jin et al. 2022

## Crowd media collage synthesizer

**Description**

In an era of citizen journalism, news feeds fill with dozens of shaky, nearly identical clips, forcing viewers to piece events together manually. To synchronize simultaneous eyewitness clips and achieve higher-quality citizen reporting, the platform can automatically detect overlap and stitch their clips into a single, synchronized multi-angle reel. On upload, computer-vision fingerprints and audio peaks cluster videos that were captured within a defined spatiotemporal window—for example, roughly 200 meters in distance and plus or minus two minutes in time. A back-end pipeline aligns frames, selects the clearest audio, and outputs either (a) a split-screen collage, (b) a swipeable angle selector, or (c) a highlight reel with seamless cuts. Each segment carries a provenance watermark—for example, “Angle 3 — @User42”—to keep its source clear. Viewers share the composite instead of dozens of near-identical shaky posts, gaining richer context and reducing selective-framing bias. Uploaders can opt out or reorder their angle before publication. Newsrooms and fact-checkers receive an API endpoint with the synchronized package, speeding verification and broadcast.

**Promise**

- One coherent reel replaces dozens of redundant single-angle uploads
- Multi-angle view exposes context, limiting selective-cropping bias
- Automatic credit tags preserve provenance and encourage participation
- API lets journalists embed verified composites within minutes
- Richer storytelling increases viewer trust and engagement

**Challenges**

- Sync errors on noisy clips (fallback to crowd alignment)
- Uploader consent concerns (simple opt-out toggle)
- High GPU cost during spikes (batch jobs, virality queue)
- Copyright or takedown conflicts (per-angle removal controls)
- Deepfake angle injection risks (hash check, authenticity flags)

**Related ideas** AI fact-checking, collaborative fact-checking, selective-scope immutable stream ledger

**References** Goode 2009

## Resolving popularity and majority bias

**Description**

On social media, popular users often get more attention simply because they already have many followers, even when others share similar or better content—a phenomenon known as *popularity bias* (Ciampaglia et al. 2018; Niitsuma et al. 2025). This creates an unfair system where new or underrepresented voices are overlooked. A similar challenge exists in academic publishing, where well-known authors or institutions may receive preferential treatment. To reduce such bias, many conferences and journals use double-blind peer review, a process in which both the authors' and reviewers' identities are concealed. This ensures that submissions are judged solely on merit. Studies have shown that double-blind review can improve evaluation fairness and raise the overall quality of selected work (Sun et al. 2022). Inspired by the double-blind review process, we propose a voluntary, fairness-driven content-ranking system that prioritizes quality over popularity. Users can opt into an anonymous evaluation process in which AI removes identifying features—such as writing style, usernames, and profile images—while preserving the core message. For visual media, if identifying features are present, they are removed or altered to conceal the creator's identity. Anonymized posts are displayed with equal visibility in a shared evaluation feed, allowing users to judge content purely on merit. After a fixed evaluation period, original posts are revealed, and their platform visibility is boosted based on performance. To further ensure fairness, the platform organizes reviewers into overlapping cultural and social communities. Content that resonates within specific groups is promoted alongside broadly appealing posts, preventing cultural majority bias. This system creates a fairer and more inclusive social media environment—one where visibility is earned through the quality of ideas, not the popularity of the creator or dominance of the majority.

**Promise**

- Fair exposure based on content quality, not user popularity
- Highlights locally valued content from different social and cultural groups
- Encourages thoughtful posting

**Challenges**

- Balancing global and local visibility is challenging
- Evaluating content is multifaceted

**Related ideas** Number-free feed, stance-based search, milestone badges for high-quality participation

**References** Ciampaglia et al. 2018, Niitsuma et al. 2025, Sun et al. 2022

# Selective-scope immutable stream ledger

**Description**

Viral eyewitness videos can be edited after the fact—key seconds trimmed, captions altered—leaving viewers and courts unsure whether the footage is genuine. Hashing every pet clip, however, would be overkill and raise privacy concerns. The *Selective-Scope Ledger* solves this dilemma by activating only when authenticity matters. It turns on when at least one of four conditions is met: (1) risky keywords such as *explosion* appear in the title, caption, or transcript; (2) a rapid-virality spike (e.g., tens of thousands of views in just a few minutes); (3) the uploader toggles an *Evidence-mode* switch; or (4) the video originates from a flagged zone or period. If none of these conditions is triggered, the upload behaves normally and creators may trim or delete it with no audit trail. When activated, the platform slices the video into several-second tiles. Each tile is signed by the device's secure chip, proving its capture time and location, and embedded with a fragile watermark that is destroyed by any later edit. Only the tiles' hashes and timestamps are written to an append-only Merkle log, batched into one root per minute to keep the ledger compact. The encrypted footage itself moves to low-cost cold storage after 30 days. Conceptually, each hash acts like a numbered seal affixed to every page of a notebook: anyone can see if a page has been removed, but no one reads the page itself unless the owner opens the book. This approach minimises overhead, preserves privacy, and still provides an immutable chain of custody for courts and journalists.

**Promise**

- Visible change records reveal exactly *what* changed and *when*
- Ordinary videos remain private, lightweight, and fast
- Public log enables journalists and courts to verify authenticity
- Pairs with the Evidence-Before-Amplify Gate to block the viral spread of tampered footage

**Challenges**

- Millions of chunks per day can slow writes and audits (subtree-optimized storage)
- Attackers could produce "authentic" hashes for inauthentic content
- A full edit history panel clogs the interface (highlight main changes)
- Keyword or spike heuristics might lock benign videos (e.g., "fire" in cooking clips) (AI context classifier cancels benign clips)

**Related ideas** AI fact-checking, crowd media collage synthesizer

**References** Crosby and Wallach 2009, McCune et al. 2010, Winkler and Rinner 2010, Fallahpour et al. 2014, Ayyalasomayajula and Ramkumar 2023

# Automated detection of filter algorithm manipulation

**Description**

When we scroll through news feeds or check online reviews, we often trust popularity signals to guide our decisions, such as likes, shares, and ratings. However, these signals can be manipulated by groups of fake accounts, like sock puppets and AI-driven bots. For example, Radivojevic et al. (2024) demonstrate that AI-driven bots can closely mimic real users and subtly spread biased messages on a large scale. How can we guard against such manipulation? Several detection algorithms have been proposed. One is *credibility scoring*, which assigns trust levels to accounts by analyzing features like account age, posting frequency, network connections, and engagement history. Another is *anomaly detection*, using models like BIRDNEST to flag suspicious reviewers whose ratings deviate from typical patterns. *Reputation thresholds* can be applied that limit the influence of new or low-trust accounts by reducing the weight of their contributions. Finally, *network analysis* can identify clusters of coordinated accounts and limit their collective impact. These algorithms work together to guard against the influence of fake accounts.

**Promise**

- Maintains crowd-driven signals
- Reduces exposure to biased content
- Protects from coordinated influence

**Challenges**

- False positives on new users
- Privacy and fairness concerns
- Scalability of detection methods
- Lack of transparency

**Related ideas** Probability ratings, discourse score, natural language feed algorithm tuning, selective-scope immutable stream ledger

**References** Ye et al. 2025, Radivojevic et al. 2024, Hooi et al. 2016, Ferrara et al. 2016, Kumar et al. 2017

## Population-representative feeds

**Description**

The algorithm selecting posts for feeds matches the distribution of political views represented in the posts to the distribution of views among all users. This mechanism prevents overrepresentation of political groups that are more active on a social media platform. The population representativeness could be enforced either at the level of all feeds or at the level of each feed. If population representativeness is enforced at the level of all feeds, individual feeds may still be biased, reflecting the user's connections and preferences, but the total number of feeds a post appears in is proportional to the portion of the user base likely to endorse it. Affinity groups of like-minded people are thus tolerated, but polarization may be reduced if moderate opinions that would otherwise be underrepresented get boosted in all groups. Alternatively, population representativeness could be enforced in each feed, exposing each user to a population-representative sample of posts. The user could also be given continuous control (0 to 1) over the degree of rebalancing toward a population-representative feed. The important concern that mere exposure to opposite views may further polarize rather than moderate political opinions (Bail 2021) may be addressed by balancing the distribution of posts at the level of all feeds and by giving the user control over the degree to which their own feed is balanced. Users could specify the reference population as either all users or users within their own national or geographically local or intellectual community.

**Promise**

- Improves accurate representation according to a democratic principle
- Exposes users to the reality of the distribution of opinions in the community
- Prevents users from mistaking the warped distribution created by social network dynamics and manipulation for the distribution of opinions

**Challenges**

- Requires a latent space of political views (nuances can be accommodated in a multidimensional latent space)
- Requires machine learning methods for placing users and posts in the latent space to match the post distribution to the user distribution
- Users might dislike a population-representative feed (users can choose to balance representativeness of the population and their followed accounts)

**Related ideas** Daily algorithm ballot, open-source forkable recommenders, discourse score

**References** Bail 2021, Stray 2021

## Graded control of thread privacy and post visibility

**Description**

A radically new element of communication on social media is that a post intended for a small audience (comparable to a remark at a dinner with friends) can be amplified to global visibility. This lends individual voices incredible potential but also renders participation in discourse risky. To enable users to control the maximum reach of their own contributions, users can be given the option to set a maximum-reach parameter by default or individually for each post. Posts with limited reach are not amplified beyond the level chosen by the user. Reach can be defined, e.g., in terms of the maximum number of shares and likes counted to determine visibility, the number of re-shares allowed, the maximum number of other users who are shown the post, the lifespan of post visibility, or a combination of these. Post visibility can also be strictly limited to specific audiences like followers, contacts-of-contacts, or thread participants only.

**Promise**

- Enables fluid control of the intimacy of social web communications
- Retains a sense of connection to a community that exceeds a small circle, while preventing unexpected and undesired blowups of visibility
- Limiting audience could make dialogue more authentic, e.g. by reducing self-censorship or performative statements
- Useful middle ground between fully-public posts and fully private content, such as DMs or posting anonymously

**Challenges**

- Bad-faith actors could still screenshot and share private or reach-limited responses (such posts should have a clearly visible indicator)
- Bad-faith actors could manipulate discourse and evade accountability by contributing posts with limited lifespan (statistics about lifespan-limited posts could help identify such actors).

**Related ideas** AI civility toolkit

**References** Gay et al. 2018, Carminati et al. 2009, Kim et al. 2025

## Reputation & community rewards

*Long-term incentives that encourage prosocial behavior*

Discourse score - *Amirhossein Farzmahdi, JohnMark Taylor*

Milestone badges for high-quality participation - *Pinyuan Feng*

AI civility toolkit - *Fan Cheng*

# Discourse score

**Description**

Likes and follower counts turn debate into a popularity contest. A *discourse score* that measures how thoughtfully someone contributes to a discussion would better align the incentives with desiderata of democratic discourse. Unlike typical platforms that reward popularity, this score highlights well-reasoned arguments. Both users and automated systems assign points for citing credible sources (such as academic papers or official documents), being followed/liked by people of diverse ideological backgrounds, and for publicly revising one's view when faced with compelling evidence. Overall, a well-designed *discourse score* could provide a guide and incentive that might transform social media into a space for open dialogue, thoughtful exchange, and genuine collaboration.

**Promise**

- Thoughtful posts gain visibility instead of viral ones
- Users earn points for compelling arguments and evidence
- A visible score helps quickly spot reliable voices

**Challenges**

- Any unidimensional operationalization might be too narrow to capture the ideal of "good discourse"
- Gaming the score: when the score becomes a target, it might cease to be a good score (Goodhart's law)
- Algorithmic bias: The score will never be completely fair and might have unintended consequences, such as over-representation of inoffensive centrist viewpoints. This might be addressed by continual adjustments of the score to factor in viewpoint diversity and accommodate more sophisticated notions of good discourse

**Related ideas** Cognicons, probability ratings, number-free feed, milestone badges for high-quality participation

**References** Wu and Zhao 2025, Hartmann et al. 2025, Irani et al. 2025, Ovadya and Thorburn 2023

## Milestone badges for high-quality participation

**Description**

One persistent problem on social media is that meaningful contributions – e.g., thoughtful debate, fact-checking, and bridging divides – are often drowned out by low-effort content and the noise of fake or low-quality accounts, making it hard for users to identify credible voices or feel motivated to participate constructively. *Milestone badges for high-quality participation* offer a positive, transparent solution: users earn visible recognition for achievements like providing well-sourced corrections, fostering civil conversations across divides, or consistently contributing high-quality summaries and syntheses. These badges not only incentivize users to engage in prosocial, constructive behaviors, but also make it easier for others to filter feeds for trustworthy participants, discouraging manipulation and fake engagement. The result is a platform where high-quality interaction is both recognized and rewarded, boosting engagement, raising discourse standards, and helping communities identify and amplify genuine contributors.

**Promise**

- More user engagement
- It can highlight credibility
- Encourages prosocial behavior
- Filters out low-quality or fake accounts

**Challenges**

- Badge inflation or gaming by encouraging quantity over quality; finding loopholes to earn badges without genuine achievement
- Lack of standards to define badges because of unclear requirements or inconsistent verification processes

**Related ideas** Cognicons, probability ratings, discourse score

**References** Anderson et al. 2013

# AI civility toolkit

**Description**

A suite of AI assistants that foster healthier discourse, protect vulnerable users, and reduce cognitive overload. Traditional moderation intervenes only after a post causes harm. This toolkit intervenes before harm occurs. Before posting, an LLM-based reframer suggests kinder wording, requests evidence, or highlights potential misunderstandings. A vulnerability shield runs in parallel: a manipulation-detector flags sycophantic or exploitative language that can target users with mental-health risks and, when “Guarded Mode” is on, rate-limits persuasive content for those users. On the consumption side, an affect-aware load balancer moderates intense or repetitive posts, prompting breaks. All modules are opt-in, offered in Light, Balanced, and Strong presets, and run on-device when possible to keep drafts private. The toolkit complements moderation by preventing problems upstream and supporting users’ self-regulation rather than simply suppressing speech.

| **Core modules** | **Purpose** | **Example UX** |
|---|---|---|
| Reframer | Suggest kinder, clearer wording before posting | Yellow preview, one-tap accept |
| Evidence Prompt | Nudge users to add sources for claims | “Add a link?” |
| Tone Checker | Highlight ad-hominem, slurs, sarcasm | Underscore toxic phrases in red |
| Load Balancer | Pace emotionally intense or repetitive content | Toast: “Want a breather?” |
| Civility Stats | Reflect on posting habits and reading diet | Daily: % constructive replies |
| Vulnerability Shield | Detect manipulation, enable Guarded Mode | “Safeguards: ON” |

**Promise**

- Converts heated drafts into constructive dialogue
- Shields vulnerable users from manipulative language
- Mitigates doomscrolling via adaptive pacing prompts
- Reduces moderator workload through upstream prevention
- Aids accessibility for neurodivergent and multilingual users

**Challenges**

- Overbearing nudges annoy (granular opt-outs)
- Misread tone/context (continuous auditing, feedback loops)
- Privacy worries about drafts (on-device inference)
- Cultural bias in civility norms (diverse training data)
- Adversarial prompt gaming (robustness tests)

**Related ideas** Ad-hominem detection

**References** Sheth et al. 2022, Garg et al. 2023, Jhaver et al. 2023, Sheng et al. 2023, Celadin et al. 2024, Williams et al. 2025, Dörr et al. 2025, Fassi et al. 2025, Thakkar et al. 2023

# Glossary

Improving social media for democratic discourse requires a shared language among people with different backgrounds, perspectives, and expertise. Some key concepts are listed and briefly described below.

**Message (post, tweet, toot):** Any discrete unit of information (e.g., an opinion, insight, news update) that a user broadcasts in an instant. These messages serve as fundamental building blocks of conversations online.

**Sharing:** The primary mechanism through which messages circulate online. It contributes to collective intelligence by distributing insights across communities.

**Popularity signals:** Indicators such as likes, shares, or star ratings that suggest how much attention or approval content has received.

**Amplification:** A system-level process where an algorithm intentionally increases the visibility of content, allowing it to reach a larger audience than possible through organic sharing alone.

**Suppression:** Opposite of amplification; involves the algorithmic de-prioritization/hiding of content to limit its visibility within users' feeds.

**Societal transparency:** The platform's openness regarding its broader impact on public discourse. This may include providing the public and researchers with the data necessary to understand how information flows and algorithmic choices affect collective democratic health.

**User transparency:** Providing users with clear, accessible information about why they are seeing specific content, and how their algorithm functions.

**Exposure diversity:** Diversity of content that users select to consume, relative to all available content.

**Nudging (for accuracy and civility)**: Using subtle interventions to guide individuals towards more accurate and civil behavior, particularly in online environments. The concept of nudging, rooted in behavioral economics, suggests that small changes in the way choices are presented can significantly influence decision-making. In the context of social media, nudging can be employed to enhance the accuracy of shared information and promote civil discourse.

**Gatekeeper's nudge**: Ranking feed content to optimize intellectual diversity.

**Model citizen nudge:** Prioritizing feed content that explicitly encourages civic virtues such as open-mindedness.

**Serendipity recommendations:** Recommendations that seek to be "surprising but relevant": Fu et al. (2023) give the four criteria of unexpectedness, novelty, diversity, and relevance. One way of providing novel viewpoints.

**Filter bubble:** Selective (or even exclusive) display of certain topics or viewpoints on a user's content feed due to operation of an algorithmic filter that omits or deprioritizes other topics or viewpoints.

**Cyberbalkanization:** Segregation of the Internet into small, intellectually homogeneous political groups with a narrow-minded approach to those with different views.

**Diversity paradox**: Concept proposed by Joris et al. (2024)—they found that while using diversity-based recommendation systems indeed exposes users to more diverse viewpoints, users did not self-report feeling as though they were reading more diversely. Authors attribute this to the invisible nature of recommendation algorithms.

**Bridging-based ranking**: Ranking functions that determine the visibility of content are bridging-based if they assign high value to content approved by an intellectually diverse set of users. Bridging-based ranking prioritizes items predicted to be valued by people who usually disagree. A bridging-based ranking function may use explicit groups or latent "perspectives" (e.g., from clustering analysis of the users). One approach is to use the product across groups of the mean group evaluations, such that only posts with positive evaluations from each group are ranked highly.

**Intelligence-based ranking:** As proposed by Burton et al. (2024), ranking content based on whether it is likely to be conducive to users forming more accurate beliefs, such as by amplifying more accurate content or users deemed more likely to be accurate based on some metrics. Another approach relies on "wisdom of the crowds": some research suggests collective accuracy often benefits from the presence of cognitively diverse individuals (such that their biases tended to "cancel out"), so an algorithm that tries to amplify a diverse range of opinions could promote collective accuracy without heavy-handedly "picking sides" on controversial issues.

**Interactional polarization**: Degree to which users come to interact mainly with those who have similar views to themselves.

**Positional polarization**: Degree to which stated political beliefs (positions) become more distinct in a population.

**Affective polarization**: Degree of expressed emotional hostility towards individuals with different political views.

**Regret**: In online learning, the commonly used objective function that measures the cumulative discrepancy between the chosen action and the optimal action. A no-regret policy is defined as one whose regret grows sublinearly in time, meaning that the policy converges to the optimal one.

**(Multi-armed) Bandit**: Common paradigm in online learning in which there is some number of levers (bandits) to pull, each with its own distribution of reward/loss. For certain scenarios, there are no-regret solutions.

**Contextual Bandit**: Scenario in which a bandit needs to be selected given a context. Used in recommender systems.

**Machine unlearning**: The task of taking a model that's been trained on a given dataset and modifying it such that for a given number of data points to "unlearn," the model performs (similarly, exactly, etc) to a model trained on the dataset minus those points. Used for privacy, safety, stale knowledge, etc.

**Reputation thresholds:** Rules that down-weight or restrict contributions from new or low-trust accounts until they've earned a baseline level of credibility.

**Sock puppets:** Fake or duplicate accounts controlled by the same actor to create false support or manipulate discussions.

**Crowdmapping:** The process of collecting, aggregating, and visualizing geolocated data submitted by a large group of individuals—often in real time—onto a digital map. It is commonly used for crisis response, citizen journalism, and community reporting, enabling users to contribute information about events or conditions in specific locations.